\documentclass[11pt]{article}
\usepackage{amsmath}
\usepackage{amsfonts}
\usepackage{amssymb}
\usepackage{graphicx}
\usepackage{pgf,tikz}
\usepackage{mathrsfs}
\usetikzlibrary{arrows}
\usepackage{float}
\usepackage{tikz}
\usetikzlibrary{decorations.pathmorphing,patterns}

\def\r{\rangle}

\def\o{\omega}

\def\r{\rangle}

\newcommand{\mbra}[1]{\langle {#1}|}
        \newcommand{\mket}[1]{ |{#1}\rangle}
        \newcommand{\sks}[2]{\langle {#1}|{#2}\rangle }
        
        \newcommand{\mop}[1]{\hat{#1}}

\author{G.Chadzitaskos\\  FNSPE,
Czech Technical University in Prague, \\B\v rehov\'a 7, CZ - 115 19 Prague,
e-mail: chadzgoc@cvut.cz}
\title{Coherent states of the asymmetric harmonic oscillator}

\begin{document}

\maketitle

\begin{abstract}
We constructed formal coherent states for an asymmetric harmonic oscillator, where the asymmetry parameter is the square root of the ratio of spring constants.  Although these states are constructed based on both Glauber's and Perelomov's approaches, in general they do not satisfy all the properties required for coherent states. Over time, the coherent states introduced in this way generally become incoherent. 
However, there are some specific parameters for the square root ratios of the spring constants $\frac{4k+1}{4l+1}$ or $\frac{4k+3}{4l+3}$. For these parameters it is possible to construct coherent states on the subspace of the Hilbert space of eigenstates. These coherent states keep their coherence during the time evolution. This case is also analyzed.
\end{abstract}

{keywords: Harmonic oscillator, Coherent states, Parabolic cylinder functions}
 
 \section{Introduction}
 A broad class of phenomena share a property of coherence.
 Coherent states influence quantum mechanics and are critical in quantum computing and quantum optics, such as lasers -- because they produce a specific phase of light. They are the closest quantum equivalent to classical harmonic motion and are important in quantum information theory. There are also quantization methods based on the construction of coherent states. A wide range of uses and applications of coherent states is described in \cite{KlauP}. In \cite{Vou} they are constructed quantum CNOT gates with input in the coherent space -- the set of all coherent spaces with logical OR and AND operations in the Boolean ring.
 
Schr\"odinger solved the problem of the time evolution of the mean value of the position operator of the quantum harmonic oscillator and introduced coherent states as the state with the same mean value time evolution as a classical oscillator coordinate \cite{Sch}.
Then in the 60s Klauder, Glauber and Sudershan used coherent states for the description of the laser light \cite{Klau,Gla1,Gla2,Sudar}.   The harmonic oscillator is the basic system used to describe black body radiation and to quantify the electromagnetic field.
Perelomov invented the group theoretical method of coherent state construction \cite{Per}.
 
A harmonic oscillator is a fundamental example of the simplest illustration of most physical laws in both classical and quantum theories. Feynman attributed the harmonic progression described by the classical harmonic oscillator equation to most processes in nature \cite{Fey}. The calculation of the spectra and eigenstates of the asymmetric harmonic oscillator is provided in \cite{Cha}.

 An asymmetric harmonic oscillator can be realised as a mass mounted on a spring, with a second spring outside or inside the first spring, not connected to the mass. 
 One for $x\geq 0$ with characteristic frequency $\omega_+$ and another for $x<0$ with characteristic frequency $\omega_-,$ as shown in Fig. 1.
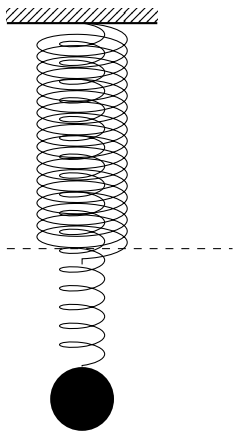
\begin{figure}
 \begin{tikzpicture}
\node[circle,fill=black,inner sep=3mm] (a) at (0,0) {};
\node[circle,inner sep=2mm] (b) at (0,1.5) {};
\draw[decoration={aspect=0.3, segment length=2.5mm, amplitude=3mm,coil},decorate] (0,5) -- (a); 
\draw[decoration={aspect=0.3, segment length=1.5mm, amplitude=6mm,coil},decorate] (0,5) -- (b); 
\fill [pattern = north east lines] (-1,5) rectangle (1,5.2);
\draw[thick] (-1,5) -- (1,5);
\draw[dashed] (-1,2) -- (2,2);
\put(2,2){\mbox{x=0}}
\end{tikzpicture} 
\caption{An asymmetric harmonic oscillator}
\end{figure}  
  
   The classical equation of motion is 
  $$\frac{d^2 x }{d t^2} + \omega^2(x)x=0,\mbox{ where }\o (x)= \lbrace \begin{smallmatrix}\o_- \hspace{.2 cm} \mbox{for} \hspace{.1cm} x<0\\  \o_+ \hspace{.2cm} \mbox{for} \hspace{.1 cm} x\geq 0 \end{smallmatrix},$$
and the Hamiltonian is
 \begin{equation}
{\cal H} = \lbrace \begin{smallmatrix} \frac{p_{x}^{2}}{2m} + \frac{1}{2}m \omega_+^{2}x^{2} \hspace{.1cm} \mbox{for} \hspace{.1cm} x \geq 0\\ \frac{p_{x}^{2}}{2m} + \frac{1}{2}m \omega_-^{2}x^{2}, \hspace{.1cm} \mbox{for} \hspace{.1cm}  x< 0 \end{smallmatrix}.
\end{equation}
The solution describes a periodic motion with the period $
 T=\frac{ \pi}{\o_-} + \frac{\pi}{\o_+}.$ 

 \section{Quantum operators}
 
 The Hamilton operator is applied separately to the positive and negative halves of the line
 
 \begin{equation}
\hat H=\frac{\hat P_{x}^{2}}{2m} + \frac{1}{2}m \omega(x)^{2}\hat{X}^{2} = \hat H_-+ \hat H_+ =  \lbrace \begin{smallmatrix} - \frac{\hbar^2}{2m}\frac{d^2}{d x^2}+\frac{1}{2}m \omega_+^{2}x^{2}, \hspace{.1cm} \mbox{for} \hspace{.1cm} x \geq 0\\ - \frac{\hbar^2}{2m}\frac{d^2}{d x^2}+\frac{1}{2}m \omega_-^{2} x^{2}, \hspace{.1cm} \mbox{for} \hspace{.1cm}  x< 0 \end{smallmatrix}.
\end{equation}
The Schr\"odinger equation is
$$ \hat H \psi (x) = E\psi (x).$$ The detailed solution of this case is described in \cite{Cha}. We will  briefly go through this.

\subsection{An asymmetric oscillator}
The standard procedure for solving this Schr\"odinger equation,
\begin{equation}{\label{Sch}}
[- \frac{\hbar^2}{2m}\frac{d^2}{d x^2}+\frac{1}{2}m \omega_\pm^{2} x^{2}] \psi(x)=E \psi(x),
\end{equation}
 is as follows:
Let's substitute into (\ref{Sch})
\begin{equation}
\xi_\pm = \sqrt{\frac{2m\omega_\pm}{\hbar}} x, \,\,\, E=\hbar \omega_\pm(\nu_\pm +\frac{1}{2}). 
\end{equation}
We get the Weber equation \cite{Weber},
\begin{equation}
[\frac{d^2}{d \xi_\pm^2}-\frac{1}{4}\xi_\pm^{2} +\nu_\pm +\frac{1}{2}] \psi(\xi_\pm)=0,
\end{equation}
 for positive axis and negative axis separately where  $\xi_- \in ( \infty,0)$ and $\xi_+ \in \langle 0, \infty).$ 

Solutions of the Schr\"odinger equation are the parabolic cylinder functions $D_{\nu_+}(\xi_+)$ for positive values of the arguments $\xi$ and $D_{\nu_-}(-\xi_-),$ and for negative arguments $\xi$. The eigenfunctions of the Hamiltonian have the form

$$
\psi(x)_{\nu}=\lbrace \begin{smallmatrix} \alpha_{\nu_ -} \, D_{\nu_-}(-\sqrt{\frac{2m \omega_-}{\hbar}} x) ,  \mbox{ for }  x<0\\  \, \alpha_{\nu_ +} \, D_{\nu_+}(\sqrt{\frac{2m \omega_+}{\hbar}} x), \hspace{.1cm} \mbox{ for } \hspace{.1cm}  x\geq 0 \end{smallmatrix}.
$$
The $\nu_-$ can be expressed as a linear function of $\nu_+$ because it must be an eigenfunction with eigenvalue $E$, so the following equations for $E,$ $\nu_+$ and $\nu_-$ are done,
\begin{eqnarray}{\label{v-v}}
E=\hbar \omega_+(\nu_+ + \frac{1}{2})=\hbar \omega_-(\nu_- + \frac{1}{2}), \, \, \nu_-=s \nu_+ + \frac{s-1}{2},
\end{eqnarray}
where $ s=\frac{\omega_+}{\omega_-}$ is the parameter that describes the oscillator.

We can assume that the parameter $s \geq 1$ without losing generality. The parameter $\nu_+$ and the coefficients $\alpha_{\nu \pm}$ are determined by the condition that the Hamiltonian must be self-adjoint. The continuity of the eigenfunction and the continuity of its derivative require the following two conditions:
 $\alpha_{\nu_-} D_{\nu_-}(0) = \alpha_{\nu_+} D_{\nu_+} (0)$ and
 $-\alpha_{\nu_-} \sqrt{\frac{2m \omega_-}{\hbar}}D'_{\nu_-} (0)  =  \alpha_{\nu_+} \sqrt{\frac{2m \omega_+}{\hbar}}D'_{\nu_+} (0),$
 which lead to the transcendental equation.
 It was solved numerically \cite{Cha}.

\subsection{Eigenvectors and eigenvalues} 

 The energy spectrum is $E_\nu = \hbar \omega_+(\nu_+ + \frac{1}{2}),$ where values $\nu_+$ are the solutions of the transcendental equation. Some numerical values of $\nu_+$ are in Tab. 1.

  \begin{table}[H]
\caption{Example of lowest values of $\nu_+$ for different parameter $s$}
{\tiny
\begin{tabular}[scale=0.2]{|l|l|l|l|l|l|l|l|l|}
\hline
$s$&$\nu_{0}$&  $\nu_{1}$& $\nu_{2}$ &$\nu_{3}$& $\nu_{4}$&$\nu_{5}$&$\nu_{6}$&$\nu_{7}$ \\
\hline
1.4& -0.0815& 0.7487&1.5841&
  2.4162&3.25019& 4.08329&4.91663&
  5.75007\\
\hline
$\sqrt{5}$ & -0.2565 & 0.1920&  0.6562&
  1.1221 &1.5858& 2.0482&2.5112& 
  2.9749\\
  \hline
4& -0.2867&0.09827& 0.4973&
  0.8995&1.3008& 1.7006&2.09984&
  2.49961\\
  \hline 
5&-0.3189& 0&0.3307& 0.6651&1&
  1.3340& 1.6672&2 \\
  \hline
$\sqrt{30}$ &-0.3309& -0.0361& 0.2695&
  0.5789& 0.8890& 1.1988&1.5077 &1.8161\\
  \hline
  \end{tabular}}
\end{table}   
One can conclude some remarks:
 \begin{itemize}
 \item The eigenfunctions depend on the ratio of the proper frequencies. 
 \item As the ratio of the proper frequencies increases, the fundamental energy level goes to zero. 
 \item Some of the eigenfunctions are performed by glued hermit functions For some rational value of $\frac{\omega_+}{\omega_-}.$
  \end{itemize}
  
 For the construction of special class of coherent states, the last statement is important.
  
\section{Fock representation}

The ratio of the frequencies $s$ is given by the spring constants. The corresponding eigenvalues $E(\nu_+) = \hbar \omega_+ (\nu_+ +\frac{1}{2})$ can be labelled by integers $n=0, 1, 2, \ldots,$ $\nu_n = \nu_{n+}$ in ascending order. The number $n$ also indicates the number of zeros in the eigenfunction. Eigenfunctions can be written using the Heaviside step function $\theta(x),$
$$\psi(x)_{E(\nu_n)} =\theta(x)(D_{\nu_{-n}}(0) D_{\nu_{+n}}(\sqrt{2 \omega_+}x))+ \theta(-x) (D_{\nu_{+n}}(0) D_{\nu_{-n}}(-\sqrt{2 \omega_-} x)),$$
for the following numerical calculation. To normalise the eigenfunctions, we can divide them by their norm,
$$\mket{s, n}  = \frac{\psi(x)_{E(\nu_n)}}{\sqrt {\int_{- \infty}^{\infty}|\psi(x)_{E(\nu_n)}|^2 dx}}.$$

The frequencies $\o_+$ and $\o_-$ are fixed by the arrangement. Then, using Fock notation, we can formally define the creation and annihilation operators as weighted shift operators -- in the same way as for the symmetric harmonic oscillator,
 $$\hat{a} |s,n\r = \sqrt{n} |s,n-1\r, \, \, \,
\mbox{ and }\,\hat{a}^+ |s,n\r = \sqrt{n+1} |s, n+1\r, $$
with $\hat{a} |s,0\r = 0.$
The annihilation and creation operators cannot be expressed in terms of position and momentum operators as they are in the symmetric case,
The operator Lie algebra is the same as for symmetric case, $$[\hat{a}, \hat{a}^+] = \hat{I}, \, \,\hat{a}^+ \hat{a} =\hat{n}, \,\,
[\hat{n},\hat{a}^+] = + \hat{a}^+,\, \,[\hat{n},\hat{a}] = - \hat{a},$$
i.e. Lie algebra $\mathsf{h}_4.$ The representative space of  $\mathsf{h}_4$ is the Hilbert space ( Fock space) $\{\mket{s,0},\mket{s,1},\mket{s,2},\mket{s,3}, \ldots \},$
$$\mop{n}\mket{s,n}=n\mket{s,n}, \,\, \mket{s,n}= \frac{(\mop{a}^+)^n}{\sqrt{n!}} \mket{s,0}. $$

\section{Coherent states}
Generalised coherent states (Glauber states) can be defined in a formal way as follows
 $$\alpha \mapsto \mket{s, \alpha}= e^{- |\alpha|^2 / 2 }
\sum^\infty_{n=0} \frac{\alpha^n }{\sqrt{n!}} \mket{s,n}.$$ Two examples of these coherent states are shown in Figures 2 and 3. 
\begin{figure}[h]
  \includegraphics[scale=0.5]{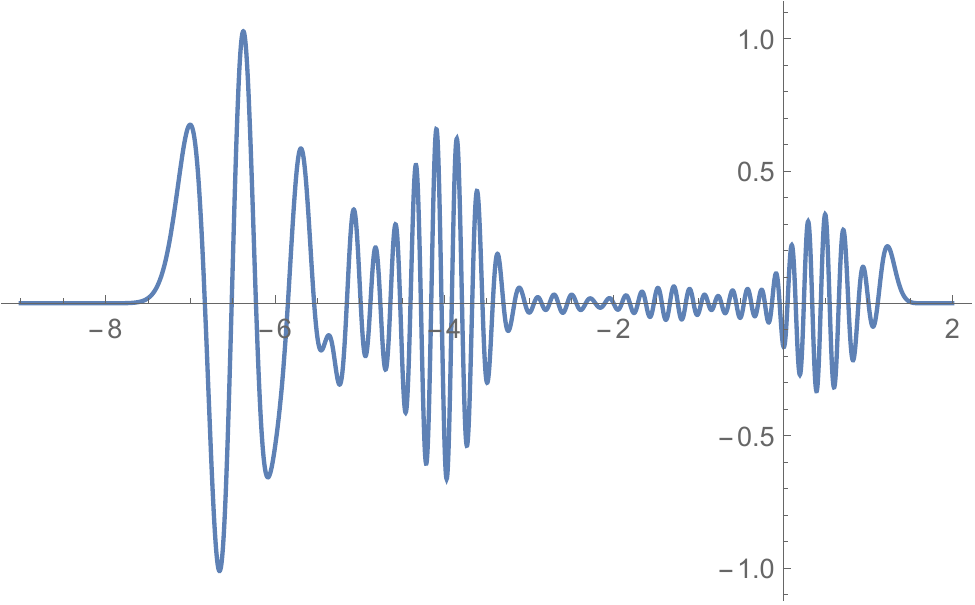}
  \caption{Coherent state $|\sqrt{26}, \alpha=8\r$, calculated with the Mathematica.}
\end{figure}

\begin{figure}[h]
\includegraphics[scale=0.5]{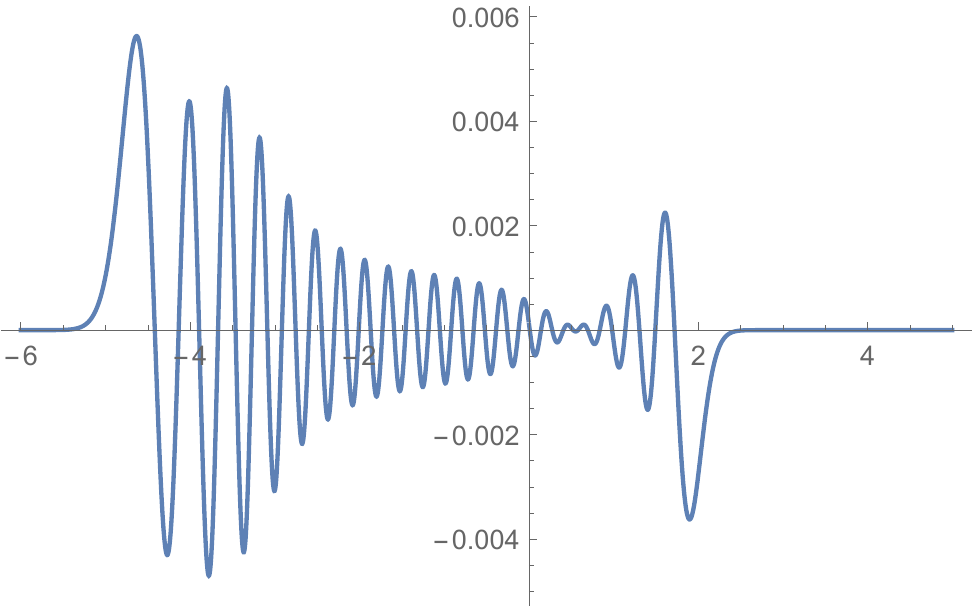} 
 \caption{Coherent state $|s=7/3,\alpha=8\r$ calculated with the Mathematica.}
\end{figure}

\subsection{Properties of coherent states}
The properties of these coherent states differ from generalised coherent states and are as follows
\begin{enumerate}
\item $ \mop{a} \mket{s,\alpha} =  \alpha  \mket{s,\alpha}, \,\, \sks{s,\alpha}{s,\alpha}= 1,  $
\item Coherent states can be created by the action of Weyl--Heisenberg group on the basic state $|s,0\r$.
Coherent states are its orbits,
$$ \mket{s,\alpha}= \exp (\alpha \mop{a}^+ - \alpha^* \mop{a}) \mket{s,0}, $$
\item The overlapping holds,
$$ \sks{s,\alpha}{s,\beta}= e^{- \frac{|\alpha|^2 + |\beta|^2 }{2} + \alpha^*  \beta} \neq 0,$$
\item The identity resolution  is also valid,
$$ \mop{I}= \int \mket{s,\alpha} \mbra{s,\alpha} \frac{\textrm{d}^2 \alpha}{\pi}$$
\item Coherent states are not Gaussian states
\item Coherent states are not coherent states in the Schr\"odinger sense
\item The time evolution does not preserve the coherence of states.
\end{enumerate}

\subsection{An illustrative example of coherent states in subspace of the Hilbert space}

Let us start with an example for the simplest special case when $s=5.$. 
Among the eigenstates there are eigenstates with equidistant energy levels. These eigenstates are chosen as the basis of the subspace, 
$$\mket{5, \tilde{n}}_s = \mket{5, 3\tilde{n}+1},\,\,\,\tilde{n}=0,1,2,\ldots.$$ 
The corresponding eigenfunctions are glued Hermite functions.
We define the creation and annihilation operators by the action on this subspace as follows:
 $$\hat{A} |5,\tilde{n}\r_s = \sqrt{\tilde{n}} |5,\tilde{n}-1\r_s, \, \, \,
\mbox{ and }\,\hat{A}^+ |5,\tilde{n}\r_s = \sqrt{\tilde{n}+1} |5, \tilde{n}+1\r_s, $$
where $\hat{A} |5,0\r_s = 0.$

The operator Lie algebra is also $\mathsf{h}_4$, i.e.
  $$[\hat{A}, \hat{A}^+] = \hat{I}, \, \,\hat{A}^+ \hat{A} =\hat{N}, \,\,
[\hat{N},\hat{A}^+] = + \hat{A}^+,\, \,[\hat{N},\hat{A}] = - \hat{A},$$
and
$$\hat{H} |5,\tilde{n}\r_s = \hbar \o_+ (\tilde{n}+1/2) |5,\tilde{n}\r_s.$$

Coherent states are defined:
$$\alpha \mapsto \mket{5, \alpha}_s= e^{- |\alpha|^2 / 2 }
\sum^\infty_{\tilde{n}=0} \frac{\alpha^{\tilde{n}} }{\sqrt{\tilde{n}!}} \mket{5,\tilde{n}}_s.$$ 
The time evolution of coherent states $|5,\alpha \r_s$ preserve the coherence,
$$\mket{s,\alpha_t}_s = e^{- \frac{\imath}{ \hbar} \mop{H} t}
\mket{s,\alpha_0}_s =   e^{- \frac{\imath \omega_+}{2} t}
\mket{s,\alpha_0 e^{- \imath \omega_+ t}}_s.$$ 
A similar conclusion can be drawn in the case of coherent states $|7/3,\alpha \r_s,$ 
$$\mket{s,\alpha_t}_s = e^{- \frac{\imath}{ \hbar} \mop{H} t}
\mket{s,\alpha_0}_s =   e^{- \frac{\imath 3 \omega_+}{2} t}
\mket{s,\alpha_0 e^{- \imath 3 \omega_+ t}}_s.$$  
  
Examples of such coherent states are shown in Fig. 4 and in the Appendix.
 
\begin{figure}[h]
  \includegraphics[scale=0.5]{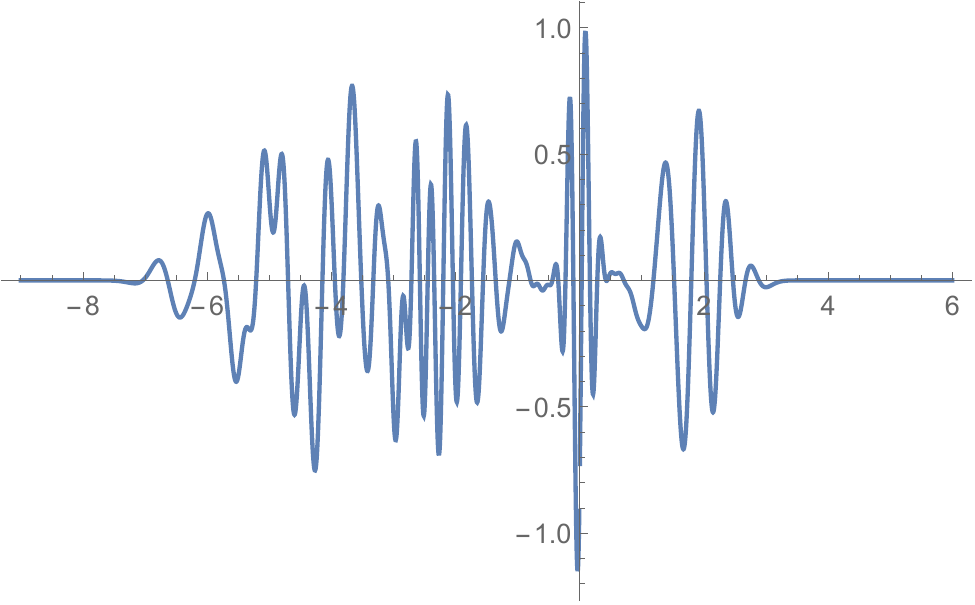}
  \caption{Coherent states $|7/3,3\r_s$  on subspace with equidistant energy levels.}
\end{figure}

$$\mket{5,\alpha_t}_s = e^{- \frac{\imath}{ \hbar} \mop{H} t}
\mket{5,\alpha_0}_s =   e^{- \frac{\imath 3 \omega_+}{2} t}
\mket{5,\alpha_0 e^{- \imath 3 \omega_+ t}}_s.$$

\subsection{General formulation of the coherent states in the sub-space of the Hilbert space}
The basic fact we will use is that for integer values of $\nu_+$ and $\nu_-$ the parabolic cylinder functions are Hermite functions, and the energy levels between such eigenstates are equidistant.
It is therefore sufficient for us to investigate only the cases where the eigenfunctions are a union of Hermite functions.

 Let us denote these integer values $n_+$ and $n_-.$
Eigenfunctions for equidistant energy levels
means that we  have to glue Hermite functions for $x<0$ and $x\ge 0,$
$$
E_k=E(n_+ (k) )=E(n_-(k) )= (n_+(k) + \frac{1}{2})\hbar\omega_+ =(n_-(k)) +
\frac{1}{2})\hbar\omega_- ,$$
where we have introduced the integer parameter $k.$
The corresponding eigenfunctions 
$|\psi_k(x)\r $ of $H$ are
\begin{equation}
|\psi_k (x)\r =  {\Huge{\lbrace}} \begin{matrix} \alpha_{n_-} H_{n_-} (\sqrt{\frac{m \omega_-}{\hbar}} x),\hspace{.1 cm} x<0\\ \alpha_{n_+} H_{n_+} (\sqrt{\frac{m \omega_+}{\hbar}} x),\hspace{.1 cm} x\geq 0, \end{matrix}
\end{equation}
where $n_+ $ and $n_- $ must be expressed in terms of the integer $k,$
where $\alpha_{n_+}$ and $\alpha_{n_-}$ are numbers to guarantee the continuity and normalisation of the eigenfunctions, and they depend on $\o_+,\o_-,n_+,n_-.$ They are
\begin{equation}
|\psi_k (x)\r =  {\Huge{\lbrace}} \begin{matrix} |\psi_{n_{-}(k)} (x)\r,\hspace{.2 cm} x<0\\ |\psi_{n_{+}(k) }(x)\r,\hspace{.2 cm} x\geq 0 \end{matrix}.
\end{equation}

Selection rules for parameters  $s$ are done by requirements:

\begin{itemize}

\item The requirement of continuity of the derivative of the eigenfunctions makes it necessary to distinguish two cases: $n_+$ and $n_-$ both take odd values or both take even values.

\item The  requirement  that $E_k$ has to be eigenvalue gives the condition that the ratio of frequencies
 $$s=\frac{\o_+}{\o_-}= \frac{(2n_- (k)+1)}{(2n_+ (k)+ 1)}=\frac{p}{q}$$ 
 must be rational number, i.e.  $p$ and $q$ must be odd. Without loosing the generality we can consider $p>q.$ 
 
\end{itemize}

Because
$$(2n_-(k) +1) = \frac{p}{q}(2n_+(k) + 1)$$ and odd integers $p$ and $q$ have no common divisor,  $2n_+(k) + 1$ takes values $q, 3q, 5q \ldots, (2l+1)\ldots $ for non negative integer $l$.
Thus we have
$$ n_+(k)= kq + \frac{q-1}{2}, \hspace{.1 cm}  n_-(k)= kp + \frac{p-1}{2},\,\,k=0,1,2,\ldots. $$
Taking in consideration $n_+(k),$ we have to chose subspace spanned on eigenfunctions corresponding to eigenvalues $E_k=(n_+(k) + \frac{1}{2})\hbar\omega_+.$ 
A minimal energy quantum for transition between two states is $kq \o_+.$

Restrict to the chosen subspace with basis vectors $|s,k\r_s = |s,\nu_+ = n_+(k)\r$, i.e.   $$ |s,0\r_s= |s,\frac{q-1}{2}\r,\,\,|s,k\r_s= |n_+(k)= kq + \frac{q-1}{2}\r$$
we can define coherent states on the subspace
$$\alpha \mapsto \mket{s, \alpha}_s= e^{- |\alpha|^2 / 2 }
\sum^\infty_{n=0} \frac{\alpha^k }{\sqrt{k!}} \mket{s,k}_s.$$ 
These coherent states preserve coherency during time evolution. 
$$\mket{s,\alpha_t}_s = e^{- \frac{\imath}{ \hbar} \mop{H} t}
\mket{s,\alpha_0}_s =   e^{- \frac{\imath q \omega_+}{2} t}
\mket{s,\alpha_0 e^{- \imath q \omega_+ t}}_s.$$

\section{Conclusion}

We have formally introduced generalised coherent states for an asymmetric harmonic oscillator. Since in general the energy levels are not equidistant, we found special cases where there are eigenstates with equidistant energy levels. We constructed the coherent states on these eigenstates.

\section*{Appendix}
Some examples of coherent states $|5,\alpha \r$  calculated by Mathematica are given in the Appendix. 

\begin{figure}[h]
  \includegraphics[scale=0.5]{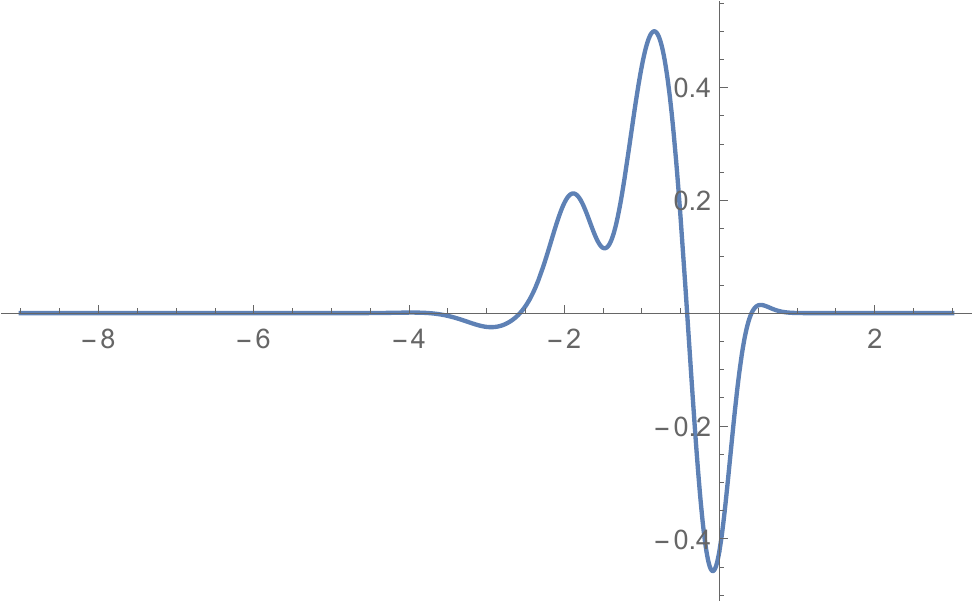}
\caption{$|5,0.5\r_s$}  
  \end{figure}

\begin{figure}[h]
  \includegraphics[scale=0.5]{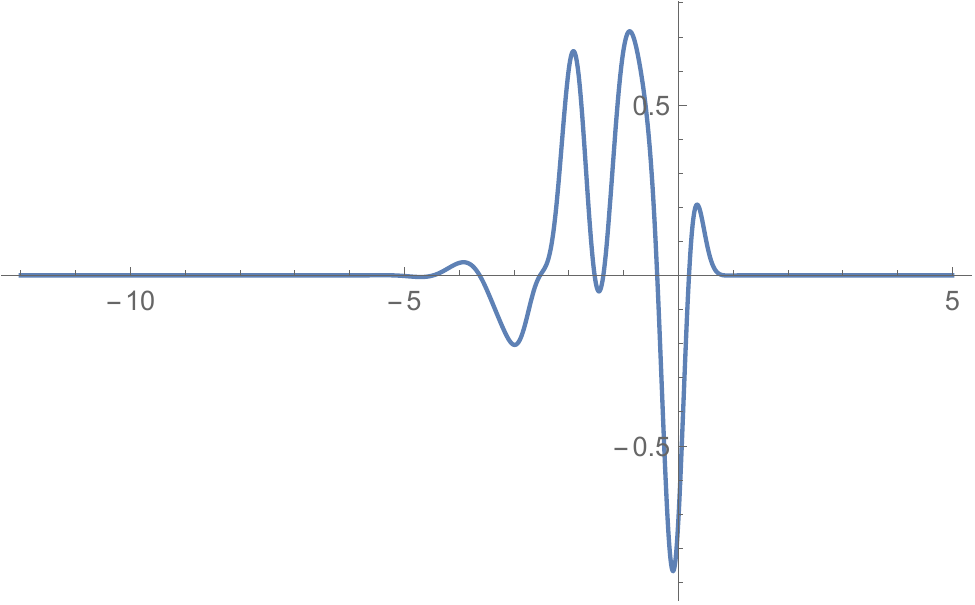}
  \caption{$|5,1\r_s$}
  \end{figure}

\begin{figure}[h]
  \includegraphics[scale=0.5]{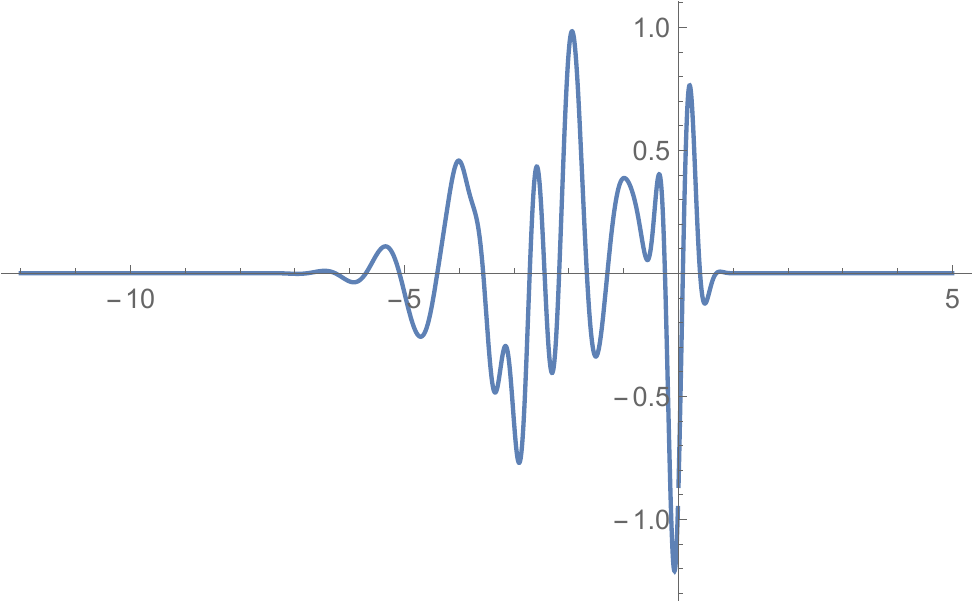}
  \caption{$|5,2\r_s$}
  \end{figure}

\begin{figure}[h]
  \includegraphics[scale=0.5]{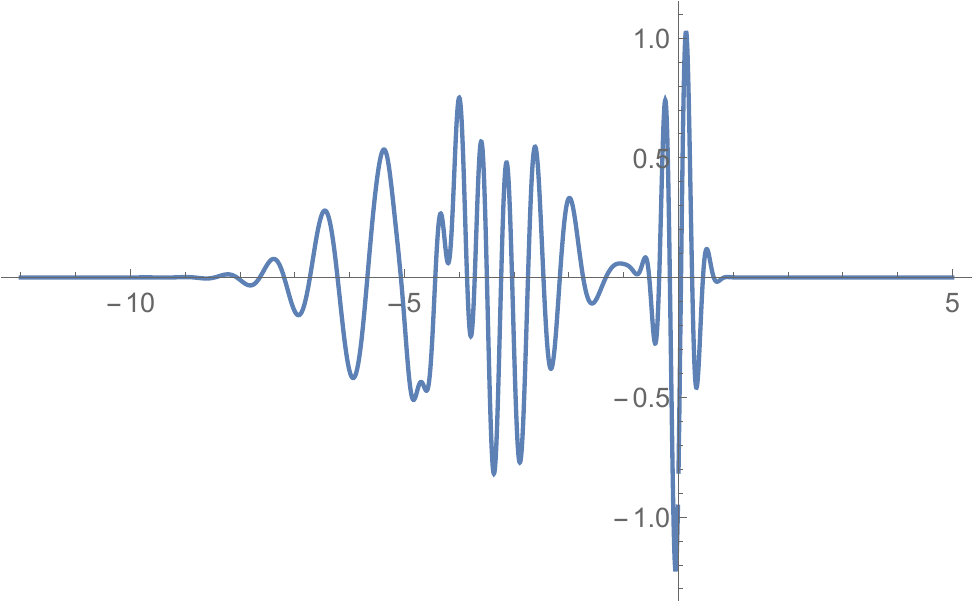}
\caption{$|5,3\r_s$}
  \end{figure}

\begin{figure}[h]
  \includegraphics[scale=0.5]{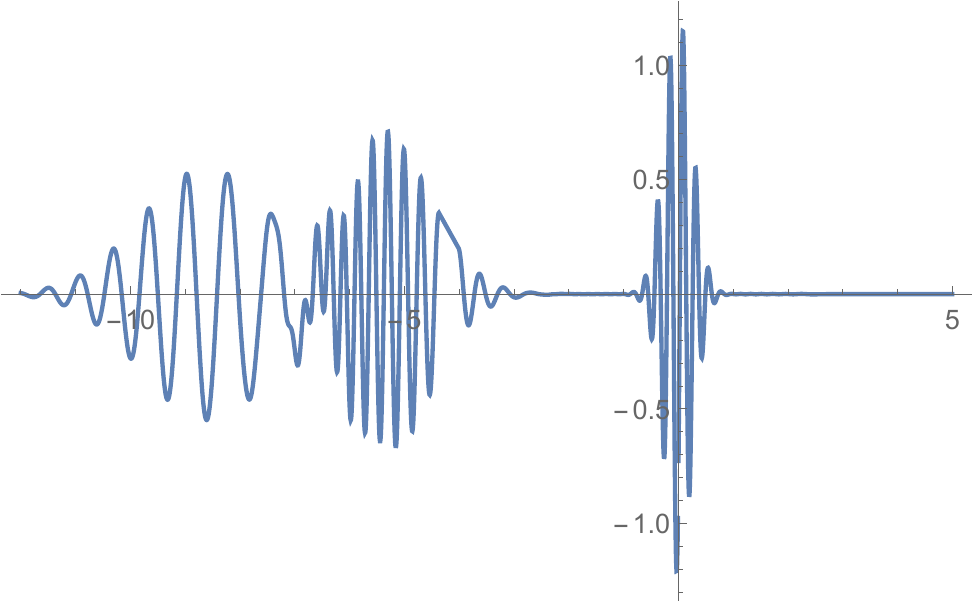}
\caption{$|5,5\r_s$}
  \end{figure}

\begin{figure}[h]
  \includegraphics[scale=0.5]{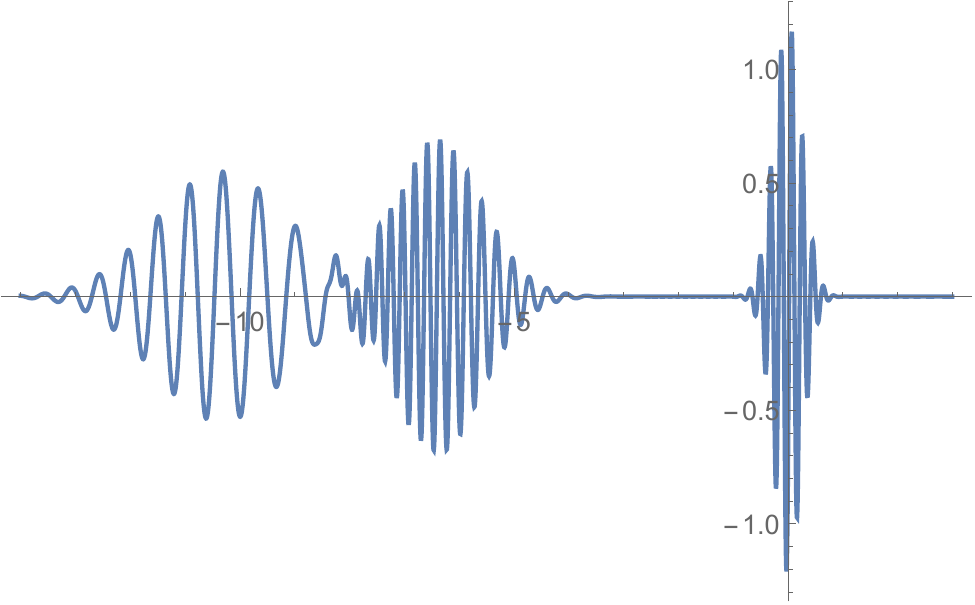}
\caption{$|5,6\r_s$}
  \end{figure}

\begin{figure}[h]
  \includegraphics[scale=0.5]{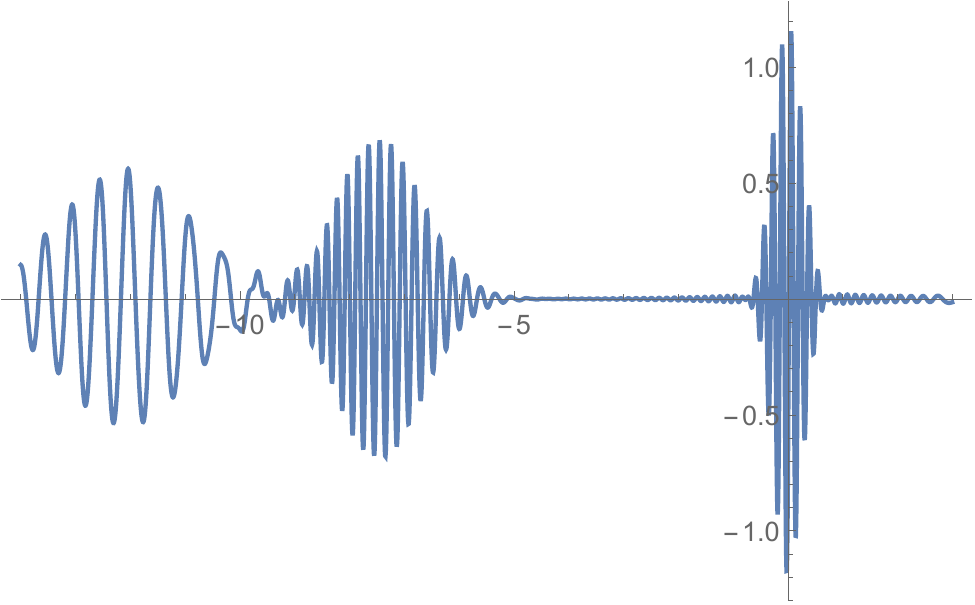}
\caption{$|5,7\r_s$}
  \end{figure}

\begin{figure}[h]
  \includegraphics[scale=0.5]{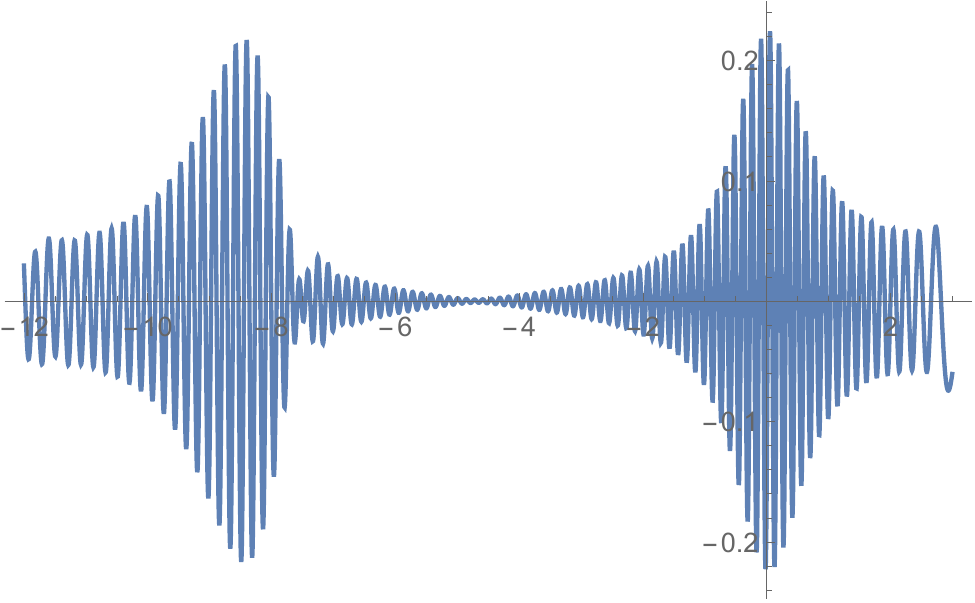}
\caption{$|5,9\r_s$}
  \end{figure}

\begin{figure}[h]
  \includegraphics[scale=0.5]{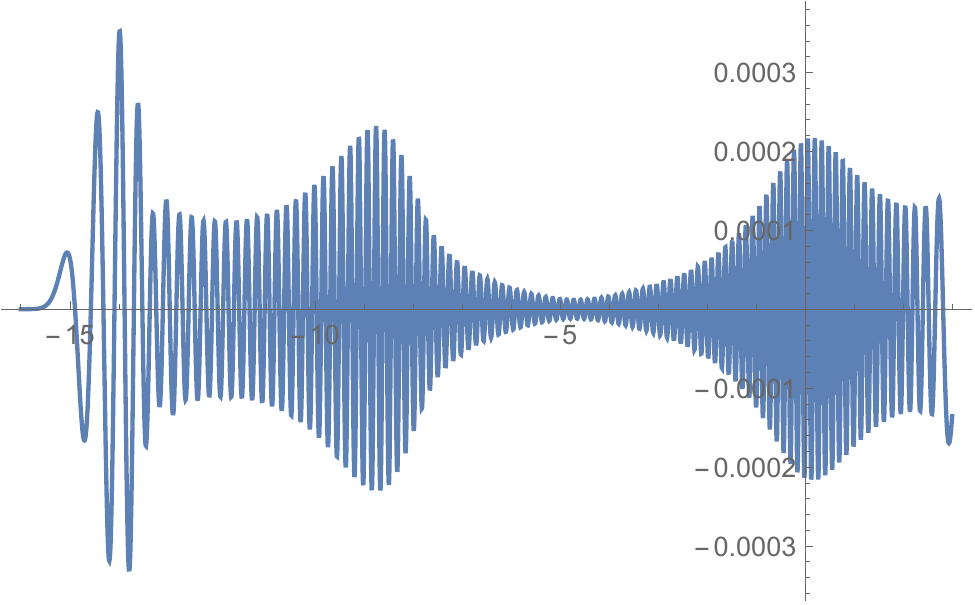}
\caption{$|5,11\r_s$}
  \end{figure}

 \end{document}